\documentclass[aps,prc,letterpaper,11pt
,twoside,tightenlines,nofootinbib,showpacs
,preprint]{revtex4-1}
\usepackage{graphicx}
\usepackage[sort&compress]{natbib}
\usepackage{subfigure}
\usepackage{amsmath}
\usepackage{amssymb}
\usepackage{amsfonts}
\usepackage{cancel}
\usepackage{hyperref}
\begin{document}

\arraycolsep1.5pt

\newcommand{\Ima}{\textrm{Im}}
\newcommand{\Rea}{\textrm{Re}}
\newcommand{\mev}{\textrm{ MeV}}
\newcommand{\be}{\begin{equation}}
\newcommand{\ee}{\end{equation}}
\newcommand{\ba}{\begin{eqnarray}}
\newcommand{\ea}{\end{eqnarray}}
\newcommand{\gev}{\textrm{ GeV}}
\newcommand{\nn}{{\nonumber}}
\newcommand{\dtres}{d^{\hspace{0.1mm} 3}\hspace{-0.5mm}}

\def\del{\partial}

\title{Wave functions of composite hadron states and relationship to couplings of scattering amplitudes for general partial waves. }

\author{F. Aceti and E. Oset}
\affiliation{
Departamento de F\'{\i}sica Te\'orica and IFIC, Centro Mixto Universidad
de Valencia-CSIC,
Institutos de Investigaci\'on de Paterna, Aptdo. 22085, 46071 Valencia,
Spain}

\date{\today}

\begin{abstract}
 
 In this paper we present the connection between scattering amplitudes in momentum space and wave functions in coordinate space, generalizing previous work done for $s$-waves to any partial wave. The relationship to the wave function of the residues of the scattering amplitudes at the pole of bound states or resonances is investigated in detail. A sum rule obtained for the couplings provides a generalization to coupled channels, any partial wave and bound or resonance states, of  Weinberg's compositeness condition, which was only valid for weakly bound states in one channel and $s$-wave. An example, requiring only experimental data, is shown for the $\rho$ meson indicating that it is not a composite particle of $\pi \pi$ and $K\bar{K}$ but something else.     

\end{abstract}

\maketitle
\tableofcontents

\section{Introduction}
\label{Intro}

  One of the most important issues in hadron spectroscopy is to determine the nature of the different hadronic states, mesons and baryons, which are reported in the PDG \cite{pdg} and still being found in new facilities like BABAR, BES, CLEO, BELLE, CERN facilities among others. The traditional wisdom that mesons are $q \bar q$ states and the baryons $qqq$ states has given room to more complicated structures in many cases, involving more quarks in one way or another   \cite{Klempt:2007cp,Klempt:2009pi}. One of the issues that has seen a more spectacular development in this field is the application of chiral dynamics to study the interaction of hadrons. The chiral Lagrangians \cite{Weinberg:1968de,Gasser:1983yg,Ecker:1994gg} represent an effective approach to QCD at low energies and use the observable mesons and baryons as degrees of freedom, which makes also comparison of prediction with experiment an easy task.
However, the perturbation theory made these Lagrangians, Chiral Perturbation theory, $\chi PT$, in spite of its spectacular success in dealing with many hadron properties, certainly could not succeed on the problem of spectroscopy, for the simple reason that resonances correspond to poles in the scattering amplitudes but poles cannot be reached in a perturbative expansion. The problem was solved with the advent of the chiral unitary approach \cite{Kaiser:1995eg,Kaiser:1996js,npa,Kaiser:1998fi,ramonet,angelskaon,ollerulf,Jido:2002yz,carmen,cola,carmenjuan,hyodo} (see \cite{review} for a review). What is found in the chiral unitary approach is that many known mesons and baryons can be explained as composite states of meson-meson or meson-baryon dynamically generated by the interaction provided by the chiral Lagrangians. An important step forward has been given in recent work on three body systems that show that also many other states can be understood as molecules of two mesons and a baryon \cite{alberto1,kanchan,xiedelta} or three mesons \cite{albertox,albertojido}.  

   One of the challenges in this field is to  determine from the experimental data the nature of the states as whether they are composite of other stable particles or something different. The pioneer in this work was Weinberg in his well known paper determining that the deuteron was a composite state (bound state from the potential) of a proton and a neutron \cite{weinberg}. More work on this issue has been done in \cite{hanhart,Hanhart:2010wh}. The generalization of Weinberg's work to coupled channels was done in \cite{juandaniel} and a different derivation is also done in \cite{jidohyodo}.
  Yet, all the work has been done for $s$-waves. The generalization of the theorem to higher partial waves is one of the contents of the present work. The theorem comes as a byproduct of a thorough study of the relationship of scattering amplitudes in momentum space to wave functions in coordinate space and the meaning of the residues at the poles, or couplings of states to the pair of interacting hadrons. The same problem was studied for just $s$-waves in \cite{juandaniel}, for bound states, and in \cite{juanjunko} for resonant states. In the present paper we generalize this work to any partial wave and both for bound states and states in the continuum. As an application of some of the results obtained we justify that the $\rho$ meson is not a composite state of $\pi \pi$ but something else. This is the commonly accepted idea about the nature of the $\rho$, which qualifies as one of the typical $q \bar q$ states, and has been supported by large $N_c$ studies done in \cite{ramon,rios,juanenrique}. The novelty is that we reach the conclusion from just the experimental data. 
  
  The work contains novel results and is rather practical concerning  applications and clarity in the derivation and in the relationship of different magnitudes. 

\section{Couplings and wave functions in the case of bound states}
\label{bs}
We want to study the nonrelativistic dynamics of a bound state generated by the interaction of two particles of masses $m_{1}$ and $m_{2}$ in a generic $l$-wave.

\subsection{Lippman-Schwinger equation}
As done in \cite{juandaniel, juanjunko} we choose as the potential V a separable function in momentum space with the modulating factor being a step function, $\Theta$. A generalization to other types of potential is done in \cite{juandaniel}, but the basic results are the same.  

Thus, our potential, this time projected in a generic $l$-wave, is 
\begin{equation}
\label{eq:potenziale}
\langle\vec{p}\ '|V|\vec{p}\rangle = V(\vec{p},\vec{p}\ ')=v(2l+1)\Theta(\Lambda-p)\Theta(\Lambda-p')P_{l}(\cos\theta)|\vec{p}|^{l}|\vec{p}\ '|^{l}\ ,
\end{equation}
where $\Lambda$ is a cutoff in the momentum space.

Let the Hamiltonian of the system be $H=H_{0}+V$, with $H_{0}$ the free Hamiltonian. The nonrelativistic Lippmann-Schwinger equation can be written as
\begin{equation}
\label{eq:lipp-schw}
T=V+V\frac{1}{E-H_{0}}T
\end{equation}
and also as
\begin{equation}
\label{eq:lipp-schw2}
T=V+V\frac{1}{E-H}V\ .
\end{equation}

Considering the second term of the series given by Eq. (\ref{eq:lipp-schw}) and substituting the expression of the potential, we can write
\begin{equation}
\label{eq:Tsecond}
\begin{split}
\langle\vec{p}|T^{(2)}|\vec{p}\ '\rangle&=\langle\vec{p}|(2l+1)v\Theta(\Lambda-p)\int_{p''<\Lambda} d^{3}p''P_{l}(\hat{p},\hat{p}'')\frac{|\vec{p}|^{l}|\vec{p}\ ''|^{l}}{E-m_{1}-m_{2}-\frac{\vec{p}\ ''^{2}}{2\mu}}\\&\times(2l+1)v\Theta(\Lambda-p')P_{l}(\hat{p}'',\hat{p}')|\vec{p}\ ''|^{l}|\vec{p}\ '|^{l}|\vec{p}\ '\rangle\ ,
\end{split}
\end{equation}
where $\mu$ is the reduced mass of the two interacting particles of masses $m_1$ and $m_2$ and we have used the normalization
\begin{equation}
\label{eq:norm}
\begin{split}
&|\vec{p}\rangle\langle\vec{p}|\equiv\int{d^{3}p}\ ,\\&
\langle\vec{p}|\vec{p}\ '\rangle=\delta^{(3)}(\vec{p}-\vec{p}\ ')\ .
\end{split}
\end{equation}
Using the expression of the Legendre functions in terms of the spherical harmonics,
\begin{equation}
\label{eq:legendre_spher}
P_{l}(\hat{p},\hat{p}'')=\frac{4\pi}{2l+1}\sum_{m}Y_{lm}(\hat{p})Y_{lm}^{*}(\hat{p}'')\ ,
\end{equation}
and their normalization condition,
\begin{equation}
\label{eq:norm_sph}
\int{d\Omega}Y_{lm}^{*}(\hat{p})Y_{l'm'}(\hat{p})=\delta_{ll'}\delta_{mm'}\ ,
\end{equation}
Eq. (\ref{eq:Tsecond}) becomes 
\begin{equation}
\label{eq:Tsecond2}
\begin{split}
T^{(2)}&=(2l+1)v\Theta(\Lambda-p)v\Theta(\Lambda-p')P_{l}(\hat{p},\hat{p}')|\vec{p}|^{l}|\vec{p}\ '|^{l}\\&\times\int_{p''<\Lambda}{d^{3}p''}\frac{|\vec{p}\ ''|^{2l}}{E-m_{1}-m_{2}-\frac{\vec{p}\ ''^{2}}{2\mu}}\ . 
\end{split}
\end{equation}
Repeating the procedure for the other terms of the expression of the Lippmann-Schwinger equation, leads to the expression of the scattering amplitude 
\begin{equation}
\begin{split}
\label{eq:ampl}
&T=(2l+1)P_{l}(\hat{p},\hat{p}')\Theta(\Lambda-p)\Theta(\Lambda-p')|\vec{p}|^{l}|\vec{p}\ '|^{l}t
\end{split}
\end{equation}
with
\begin{equation}
\label{eq:t}
t=v+v\ G\ t\ ,\ \ \ \ \ t=\frac{v}{(1-vG)}=\frac{1}{v^{-1}-G}\ ,
\end{equation}
where
\begin{equation}
\label{eq:loop}
G=\int_{_{p''<\Lambda}}{d^3p''\frac{|\vec{p}\ ''|^{2l}}{E-m_{1}-m_{2}-\frac{\vec{p}\ ''^{2}}{2\mu}}}\ .
\end{equation}
We can see that the factor $2l+1$ does not appear in the equation for $t$.

We can also see that $v$ in Eq. (\ref{eq:t}) does  not contain $|\vec{p}|^{l}$, which is now absorbed into the definition of the loop function $G$ of Eq. (\ref{eq:loop}). Other approaches for $p$-waves, like the one of \cite{palomar,michael}, factorize on shell $|\vec{p}|^{l}$ and associate it to the potential. In those cases, $G$ does not have the factor $|\vec{p}''|^{2l}$ in Eq. (\ref{eq:loop}). The procedure leads to the same $ImT$ but can induce differences in the $ReT$. Later on we shall see that the option chosen here, which stems from the form of the potential in Eq. (\ref{eq:potenziale}), allows one to generalize the sum rule for the couplings found in \cite{juandaniel}, which is lost if one uses the on shell factorized form.                                                                                                                                                                                                                                                                                                                                                                                                                                                                                                                                                                                                                                                                                                                                                                                                                                                                                                                                                                                                                                                                                                                                                                                                                                                                                                                                                                                                                                                                                                                                                                                                                                                                                                                                                                                                                                                                                                                                                                                                                                                                                                                                                                                                                                                                                                                                                                                                                                                                                                                                                                                                                                                                                                                                                                                                                                                                                                                                                                                                                                                                                                                                                                                                                                                                                                                                                                                                                                                                                                                                                                                                                                                                                                                                                                                                                                                                                                                                                                                                                                                                                                                                                                                                                                                                                                                                                                                                                                                                                                                                                                                                                                                                                                                                                                                                                                                                                                                                                                                                                                                                                                                                                                                                                                                                                                                                                                                                                                                                                                                                                                                                                                                                                                                                                                                                                                                                                                                                                                                                                                                                                                                                                                                                                                                                                                                                                                                                                                                                                                                                                                                                                                                                                                                                                                                                                                                                                                                                                                                                                                                                                                                                                                                                                                                                                                                                                                                                                                                                                                                                                                                                                                                                                                                                                                      
 
\subsection{Bound states in the one channel case: wave function in the momentum space}
From the Schr\"{o}dinger equation follows
\begin{equation}
\label{eq:wav}
|\Psi\rangle=\frac{V}{E-H_{0}}|\Psi\rangle\ ,
\end{equation}
which has the solution  
\begin{equation}
\label{eq:wav2}
\langle\vec{p}|\Psi\rangle=\int{d^{3}k}\int{d^{3}k'}\langle\vec{p}| \frac{1}{E-H_{0}}|\vec{k}\rangle\langle\vec{k}|V|\vec{k}'\rangle\langle\vec{k}'|\Psi\rangle\ .
\end{equation}
Substituting the potential (\ref{eq:potenziale}) and using
\begin{equation}
\langle\vec{p}| \frac{1}{E-H_{0}}|\vec{k}\rangle=\delta^{(3)}(\vec{p}-\vec{k})\frac{1}{E-m_{1}-m_{2}-\frac{\vec{p}^2}{2\mu}}\ ,
\end{equation}
we obtain 
\begin{equation}
\label{eq:wav3}
\begin{split}
\langle\vec{p}|\Psi\rangle&=\int_{k'<\Lambda}{d^{3}k'}(2l+1)v\Theta(\Lambda-p)\frac{1}{E-m_{1}-m_{2}-\frac{\vec{p}^2}{2\mu}}|\vec{p}|^{l}|\vec{k}'|^{l}P_{l}(\hat{p},\hat{k}')\langle\vec{k}'|\Psi\rangle\\&=4\pi\sum_{m}\frac{\Theta(\Lambda-p)|\vec{p}|^{l}v}{E-m_{1}-m_{2}-\frac{\vec{p}^2}{2\mu}}Y_{lm}(\hat{p})\int_{k<\Lambda}{d^{3}k}Y_{lm}^{*}(\hat{k})|\vec{k}|^{l}\langle\vec{k}|\Psi\rangle\ ,
\end{split}
\end{equation}
which gives us the expression of the wave function in momentum space.

Defining $\langle\vec{k}|\tilde{\Psi}\rangle$ as
\begin{equation}
\label{eq:lincomb}
\langle\vec{k}|\Psi\rangle\cong(4\pi)^{1/2}\sum_{m'}a_{m'}Y_{lm'}(\hat{k})\langle\vec{k}|\tilde{\Psi}\rangle\ ,
\end{equation}
with the coefficients $a_{m'}$ normalized as
\begin{equation}
\label{eq:a_m}
\sum_{m'}|a_{m'}|^2=1\ ,
\end{equation}
we can write Eq. (\ref{eq:wav3}) as
\begin{equation}
\label{eq:wav4}
\begin{split}
\langle\vec{p}|\Psi\rangle&=(4\pi)^{1/2}\sum_{m}a_{m}Y_{lm}(\hat{p})\langle\vec{p}|\tilde{\Psi}\rangle\\&=(4\pi)^{1/2}\sum_{m}\frac{\Theta(\Lambda-p)|\vec{p}|^{l}v}{E-m_{1}-m_{2}-\frac{\vec{p}^2}{2\mu}}a_{m}Y_{lm}(\hat{p})\int_{k<\Lambda}{d^{3}k}|\vec{k}|^{l}\langle\vec{k}|\tilde{\Psi}\rangle
\end{split}
\end{equation}
obtaining
\begin{equation}
\label{eq:wav5}
\begin{split}
\langle\vec{p}|\tilde{\Psi}\rangle=\frac{\Theta(\Lambda-p)|\vec{p}|^{l}v}{E-m_{1}-m_{2}-\frac{\vec{p}^2}{2\mu}}\int_{k<\Lambda}{d^{3}k}|\vec{k}|^{l}\langle\vec{k}|\tilde{\Psi}\rangle\ .
\end{split}
\end{equation}
 
Integrating in $d^{3}p$ and multiplying both sides by $|\vec{p}|^{l}$, Eq. (\ref{eq:wav5}) becomes 
\begin{equation}
\label{eq:pole}
\begin{split}
\int{d^{3}p}|\vec{p}|^{l}\langle\vec{p}|\tilde{\Psi}\rangle&=\int_{p<\Lambda}{d^{3}p}\frac{|\vec{p}|^{2l}v}{E-m_{1}-m_{2}-\frac{\vec{p}^{2}}{2\mu}}\int_{k<\Lambda}{d^{3}k}\langle\vec{k}|\tilde{\Psi}\rangle|\vec{k}|^{l}\\&=G\ v\ \int_{k<\Lambda}{d^{3}k}\langle\vec{k}|\tilde{\Psi}\rangle|\vec{k}|^{l}\ ,
\end{split}
\end{equation}
giving us the condition for a pole in t-matrix corresponding to a bound state,
\begin{equation}
\label{eq:pole2}
1-G(E)\ v=0\ ,
\end{equation}
which will occur for some value $E_{\alpha}<m_1+m_2$.
 
\subsection{Normalization of the wave function}
Let $E_{\alpha}<m_1+m_2$ be the solution of Eq. (\ref{eq:pole2}). Since we are dealing with a bound state, its wave function will satisfy the normalization condition
\begin{equation}
\label{eq:prob}
\int{d^{3}p}|\langle\vec{p}|\Psi\rangle|^{2}=1\ .
\end{equation}
We can now substitute the expression of the wave function (\ref{eq:wav4}) in the above equation,
\begin{equation}
\label{eq:prob2}
\begin{split}
\int{d^{3}p}|\langle\vec{p}|\Psi\rangle|^{2}&=\int{d^{3}p}(4\pi)^{1/2}\sum_{m}\frac{\Theta(\Lambda-p)|\vec{p}|^{l}v}{E-m_1-m_2-\frac{\vec{p}^2}{2\mu}}\ a_{m}^{*}Y_{lm}^{*}(\hat{p})\int_{k<\Lambda}{d^{3}k}\langle\tilde{\Psi}|\vec{k}\rangle|\vec{k}|^{l}\\&\times(4\pi)^{1/2}\sum_{m'}\frac{\Theta(\Lambda-p)|\vec{p}|^{l}v}{E-m_1-m_2-\frac{\vec{p}^2}{2\mu}}\ a_{m'}Y_{lm'}(\hat{p})\int_{k'<\Lambda}{d^{3}k'}\langle\vec{k}'|\tilde{\Psi}\rangle|\vec{k}'|^{l}\\&=\int_{p<\Lambda}{d^{3}p}\left(\frac{|\vec{p}|^{l}v}{E-m_1-m_2-\frac{\vec{p}^2}{2\mu}}\right)^{2}\sum_{m}|a_{m}|^{2}\left|\int_{k<\Lambda}{d^{3}k|\vec{k}|^{l}\langle\vec{k}|\tilde{\Psi}\rangle}\right|^{2}\ .
\end{split}
\end{equation}
 
Taking into account the normalization in Eq. (\ref{eq:a_m}) and Eq. (\ref{eq:loop}), we obtain
\begin{equation}
\label{eq:prob3}
-\frac{dG}{dE}\ v^{2}\left|\int_{k<\Lambda}{d^{3}k}|\vec{k}|^{l}\langle\vec{k}|\tilde{\Psi}\rangle\right|^{2}=1\ .
\end{equation}

Eq. (\ref{eq:prob3}) is relevant for our purposes. By construction, the left-hand side is the probability that the bound state found couples to the hadron-hadron component under consideration. We shall see in the following sections that, when we have several interacting hadron-hadron pairs, it is replaced by a sum over the different coupled channels, and each of the terms indicates the probability to find each hadron-hadron pair in the wave function. Yet, it could be that a physical state couples not only to hadron-hadron pairs, but also to a different component of non molecular type, say $q\bar{q}$ for mesons or $qqq$ for baryons. One example of this can be found in studies with the chiral bag model \cite{gerry} where the $\Delta$ has a big $qqq$ component and a smaller $\pi N$ one. In this case the normalization would be given by
\begin{equation}
\label{eq:gerry}
-\frac{dG}{dE}\ v^{2}\left|\int_{k<\Lambda}{d^{3}k}|\vec{k}|^{l}\langle\vec{k}|\tilde{\Psi}\rangle\right|^{2}+|\langle\beta|\Psi\rangle|^2=1\ ,
\end{equation}
and then
\begin{equation}
\label{eq:prob4}
-\frac{dG}{dE}\ v^{2}\left|\int_{k<\Lambda}{d^{3}k}|\vec{k}|^{l}\langle\vec{k}|\tilde{\Psi}\rangle\right|^{2}=1-Z\ ;\ \ \ \ \ \ \ Z=|\langle\beta|\Psi\rangle|^2\ ,
\end{equation}
where $|\beta\rangle$ is the genuine component of the state (the $qqq$ in the bag model of \cite{gerry}). 

This is the statement of the compositeness condition of Weinberg, although derived and formulated in a different way. The idea is that the first term in Eq. (\ref{eq:prob3}) represents the probability to find the hadron-hadron component in the wave function, and its diversion from unity is the probability to find some other components which have not been considered if one has neglected important channels of the more general coupled channels problem. Yet, the theorem can be stated in a more practical way as we show in the following section.

\subsection{Couplings}
Now we want to use the other form of the Lippmann-Schwinger equation,  Eq. (\ref{eq:lipp-schw2}). We have
\begin{equation}
\label{eq:lipsch}
\langle\vec{p}|T|\vec{p}\ '\rangle=\langle\vec{p}|V|\vec{p}\ '\rangle+\sum_{nn'}\langle\vec{p}|V|n\rangle\langle n|\frac{1}{E-H}|n'\rangle\langle n'|V|\vec{p}\ '\rangle\ ,
\end{equation}
where $|n\rangle$ and $|n'\rangle$ are complete sets of eigenstates of the full Hamiltonian $H$. 

In the vicinity of the pole, $E=E_{\alpha}$, we can take into account only the dominant contribution coming from the eigenstate $|\alpha\rangle$. Then,
\begin{equation}
\label{eq:lipsch2}
\langle\vec{p}|T|\vec{p}\ '\rangle\sim\langle\vec{p}|V|\vec{p}\ '\rangle+\int{d^{3}k}\int{d^{3}k'}\langle\vec{p}|V|\vec{k}\rangle\langle\vec{k}|\alpha\rangle\frac{1}{E-E_{\alpha}}\langle\alpha|\vec{k}'\rangle\langle\vec{k}'|V|\vec{p}\ '\rangle\ .
\end{equation}

We can write Eq. (\ref{eq:lipsch2}) explicitly
\begin{equation}
\label{eq:lipsch3}
\begin{split}
\langle\vec{p}|T|\vec{p}\ '\rangle&=4\pi\Theta(\Lambda-p)\Theta(\Lambda-p')t|\vec{p}|^{l}|\vec{p}\ '|^{l}\sum_{m}Y_{lm}(\hat{p})Y^{*}_{lm}(\hat{p}')\\&=4\pi\Theta(\Lambda-p)\Theta(\Lambda-p')v|\vec{p}|^{l}|\vec{p}\ '|^{l}\sum_{m}Y_{lm}(\hat{p})Y^{*}_{lm}(\hat{p}')\\&+\int_{k<\Lambda}{d^{3}k}\int_{k'<\Lambda}{d^{3}k'}4\pi\Theta(\Lambda-p)v|\vec{p}|^{l}|\vec{k}|^{l}\sum_{m'}Y_{lm'}(\hat{p})Y^{*}_{lm'}(\hat{k})\\&\times\langle\vec{k}|\alpha\rangle\frac{1}{E-E_{\alpha}}\langle\alpha|\vec{k}'\rangle4\pi\Theta(\Lambda-p')v|\vec{k}'|^{l}|\vec{p}\ '|^{l}\sum_{m''}Y_{lm''}(\hat{k}')Y^{*}_{lm''}(\hat{p}')\ .
\end{split}
\end{equation}

Using the fact that 
\begin{equation}
\langle\vec{k}|\alpha\rangle\frac{1}{E-E_{\alpha}}\langle\alpha|\vec{k}'\rangle\rightarrow 4\pi\frac{1}{E-E_{\alpha}}\langle\vec{k}|\tilde{\alpha}\rangle\langle\tilde{\alpha}|\vec{k}'\rangle\sum_{m}Y_{lm}(\hat{k})Y^{*}_{lm}(\hat{k}')\ ,
\end{equation}
from Eq. (\ref{eq:lipsch3}) we can easily find 
\begin{equation}
\label{eq:tmat}
t=v+v^{2}\frac{1}{E-E_{\alpha}}\left|\int_{k<\Lambda}{d^{3}k}\langle\vec{k}|\tilde{\alpha}\rangle|\vec{k}|^{l}\right|^{2}\ .
\end{equation}
Close to the pole the coupling $g$ is defined such that the amplitude can be written as
\begin{equation}
\label{eq:tmexp}
t=\frac{g^{2}}{E-E_{\alpha}}\ , 
\end{equation}
and hence
\begin{equation}
\label{eq:coup}
g^2=\lim_{E\rightarrow E_{\alpha}}(E-E_{\alpha})\ t=v^{2}\left|\int_{k<\Lambda}{d^{3}k}\langle\vec{k}|\tilde{\alpha}\rangle|\vec{k}|^{l}\right|^{2}\ . 
\end{equation}
Eq. (\ref{eq:coup}) allows to write $v^{2}\left|\int_{k<\Lambda}{d^{3}k}\langle\vec{k}|\tilde{\alpha}\rangle|\vec{k}|^{l}\right|^{2}$ in terms of the couplings, which can be determined experimentally, and then Eq. (\ref{eq:prob3}) can be stated, for a composite state, as 
\begin{equation}
\label{eq:coup2}
-g^2\frac{dG}{dE}=1 
\end{equation} 
and, in general, as
\begin{equation}
\label{eq:coup3}
-g^2\frac{dG}{dE}=1-Z\ , 
\end{equation} 
in the case we have coupling to a genuine component, with $Z$ the probability to find this component in the wave function. Since $g$ can be determined experimentally, one can see the value of Eq. (\ref{eq:coup3}) to determine the nature of the states, as has been made manifest in \cite{weinberg, hanhart}.

We have been assuming implicitly that $v$ is energy independent. Indeed, Eq. (\ref{eq:coup2}) can be obtained from Eq. (\ref{eq:t}) using l'H\^{o}pital rule
\begin{equation}
\label{eq:l'hopital}
g^2=\lim_{E\to E_{\alpha}}(E-E_{\alpha})t=\lim_{E\to E_{\alpha}}\frac{E-E_{\alpha}}{v^{-1}-G}=\frac{1}{-\frac{dG}{dE}}\ ,
\end{equation}
where in the last step we have assumed that $v$ is energy independent.
Here we see the convenience of avoiding the on shell factorization, where the vertex $|\vec{p}|^{2l}$ is incorporated in $v$, because then the new $v$ is necessarily energy dependent and we cannot obtain Eq. (\ref{eq:l'hopital}). 

Actually, as seen in \cite{weinberg, hanhart}, $Z$ means the probability to have the genuine component of the state. When dealing with a physical system in which $Z\neq 0$, one can accomodate it in the present formalism by taking a potential $v$ energy independent, which accounts for the couplings to the hadron-hadron component, and a $CDD$ pole term \cite{castillejo} of the type $a/(E-E_R)$, which accounts for the coupling to the genuine component. As shown in \cite{nsd}, this is a good tool for the analysis of data that return $Z\leq 1$ as it should be, with $Z$ related to the strenght of the $CDD$ pole, $a$.

\subsection{Wave function in coordinate space}
We can evaluate the wave function in coordinate space as
\begin{equation}
\begin{split}
\label{eq:wfcoor}
\langle\vec{x}|\Psi\rangle&=\int{d^{3}p}\langle\vec{x}|\vec{p}\rangle\langle\vec{p}|\Psi\rangle=\int{\frac{d^{3}p}{(2\pi)^{2/3}}}e^{i\vec{p}\vec{x}}\langle\vec{p}|\Psi\rangle\\&=\int_{p<\Lambda}{\frac{d^{3}p}{(2\pi)^{2/3}}}e^{i\vec{p}\vec{x}}\frac{|\vec{p}|^{l}}{E-m_1-m_2-\frac{\vec{p}^2}{2\mu}}g(4\pi)^{1/2}\sum_{m}a_{m}Y_{lm}(\hat{p})\ ,
\end{split}
\end{equation}
where we have used Eqs. (\ref{eq:wav4}) and  (\ref{eq:coup2}) in the expression of the wave function in momentum space.

Recalling the expansion of the plane wave in terms of spherical harmonics and Bessel functions
\begin{equation}
\label{eq:planewave}
e^{i\vec{p}\vec{x}}=4\pi\sum_{lm}i^{l}j_{l}(pr)Y^{*}_{lm}(\hat{p})Y_{lm}(\hat{r})\ ,
\end{equation}
Eq. (\ref{eq:wfcoor}) becomes
\begin{equation}
\label{eq:wfcoor2}
\begin{split}
\langle\vec{x}|\Psi\rangle=g\ \int_{p<\Lambda}{\frac{d^{3}p}{(2\pi)^{2/3}}}\ i^{l}j_{l}(pr)\frac{|\vec{p}|^{l}}{E-m_1-m_2-\frac{\vec{p}^2}{2\mu}}(4\pi)^{1/2}\sum_{m}a_{m}Y_{lm}(\hat{r})\ , 
\end{split}
\end{equation}
which gives us the expression of the wave function in coordinate space.

The Bessel functions satisfy
\begin{equation}
\label{eq:bess_inf}
j_{l}(z)\overset{z\to\infty}{\longrightarrow} \frac{1}{z}\cos\left[z-\frac{l+1}{2}\pi\right]\ .     
\end{equation}
Thus, the asymptotic behaviour of the wave function is
\begin{equation}
\label{eq:coor_infty}
\langle\vec{x}|\Psi\rangle\overset{r\to\infty}{\longrightarrow}-g\ 2\mu\ \sqrt{2}\pi\left(1+O\left(\frac{1}{\Lambda}\right)\right)\sum_{m}a_{m}Y_{lm}(\hat{r})(i\gamma)^{l}\ \frac{e^{-\gamma r}}{r}\ ,
\end{equation}
where $\gamma=\sqrt{2\mu|E_B|}$, with $E_B$ the binding energy, $E_B=E-m_1-m_2<0$.

\subsection{The meaning of the couplings in terms of wave functions}
Now we want to establish the relation between the coupling and the wave function at the origin in coordinate space.

From the behaviour of the Bessel functions for small values of the argument,
\begin{equation}
\label{eq:bessel}
j_{l}(pr)\longrightarrow\frac{|\vec{p}|^{l}|\vec{r}|^{l}}{(2l+1)!!} \ \ \ \ \ \ \ \ \ \ \ \ |\vec{p}||\vec{r}|\rightarrow 0\ , 
\end{equation}
follows the expression of the wave function at the origin
\begin{equation}
\label{eq:wfcoordorigin}
\Psi(\vec{x}\rightarrow0)=\langle\vec{x}\rightarrow 0|\Psi\rangle=g\ G\ \frac{i^{l}|\vec{r}|^{l}}{(2\pi)^{2/3}(2l+1)!!}(4\pi)^{1/2}\sum_{m}a_{m}Y_{lm}(\hat{r})\ .
\end{equation}
Eq. (\ref{eq:wfcoordorigin}) leads to the relation between the coupling and the wave function
\begin{equation}
\label{eq:coupwf}
g= G^{-1}\ \hat{\Psi}\ , 
\end{equation}
where we have defined
\begin{equation}
\hat{\Psi}=\frac{(2\pi)^{2/3}(2l+1)!!}{i^{l}|\vec{r}|^{l}(4\pi)^{1/2}\sum_{m}a_{m}Y_{lm}(\hat{r})}\ \Psi(\vec{x}\rightarrow 0)\ .
\end{equation}
For $l=0$ this equation leads to
\begin{equation}
\label{eq:l=0}
(2\pi)^{3/2}\Psi(0)=gG\ ,
\end{equation}
which is the result obtained in  \cite{juandaniel}.

\subsection{Generalization to coupled channels}
Now we have
\begin{equation}
\label{eq:potential_cc}
\langle\vec{p}|V|\vec{p}\ '\rangle\equiv (2l+1)\ v\ \Theta(\Lambda-p)\Theta(\Lambda-p)|\vec{p}|^{l}|\vec{p}\ '|^{l}P_{l}(\cos\theta)\ , 
\end{equation}
where $v$ is a $N\times N$ matrix with $N$ the number of channels.
The expressions obtained can be generalized to many channels.

\subsection{Couplings in coupled channels}
We can write the $t$ matrix as
\begin{equation}
\label{eq:tmatrix_cc2}
t=[1-vG]^{-1}v\ ,
\end{equation}
where $G$ is the diagonal matrix 
\begin{equation}
\begin{pmatrix}
  G_1 & \  & \  & \  \\
  \  & G_2 & \  & \  \\
  \  & \  & \ddots & \ \\
  \  & \  & \  & G_N
 \end{pmatrix}
 \,
\end{equation} 
with $G_{i}$ given by Eq. (\ref{eq:loop}) for each channel.
This can be rewritten as
\begin{equation}
\label{eq:tmatrix_cc2}
t=\frac{\it{A}v}{det(1-vG)}\ , 
\end{equation}
where $\it{A}$ is defined as
\begin{equation}
\label{eq:Amatrix}
\it{A}=[det(1-vG)](1-vG)^{-1} 
\end{equation}
and is introduced to single out the source of the pole in coupled channels, which is given by the condition
\begin{equation}
\label{eq:polecondition}
det(1-vG)=0\ .
\end{equation}
 
Now we have
\begin{equation}
\label{eq:coup_cc}
g_{i}g_{j}=\lim_{E\rightarrow E_{\alpha}}(E-E_{\alpha})\ t_{ij}=\left[\frac{(\it{A}v)_{ij}}{\frac{d}{dE}det(1-vG)}\right]_{E=E_{\alpha}}\ ,
\end{equation}
and hence
\begin{equation}
\label{eq:cou_cc2}
\frac{g_{j}}{g_{i}}=\left[\frac{(\it{A}v)_{ij}}{(\it{A}v)_{ii}}\right]_{E=E_{\alpha}}\ . 
\end{equation}

\subsection{Wave functions in momentum space}
Eq. (\ref{eq:wav4}) is generalized as follows
\begin{equation}
\label{eq:wf_cc}
\langle\vec{p}|\Psi_{i}\rangle=\frac{|\vec{p}|^{l}\Theta(\Lambda-p)(2l+1)}{E-M_{i}-\frac{\vec{p}^2}{2\mu_{i}}}\sum_{j}v_{ij}\int_{k<\Lambda}{d^{3}k}|\vec{k}|^{l}P_{l}(\hat{k},\hat{p})\langle\vec{k}|\Psi_{j}\rangle\ ,
\end{equation}
and using Eq. (\ref{eq:lincomb}) we find
\begin{equation}
\label{eq:wf_cc2}
\langle\vec{p}|\tilde{\Psi}_{i}\rangle=\frac{|\vec{p}|^{l}\Theta(\Lambda-p)}{E-M_{i}-\frac{\vec{p}^2}{2\mu_{i}}}\sum_{j}v_{ij}\int_{k<\Lambda}{d^{3}k}|\vec{k}|^{l}\langle\vec{k}|\tilde{\Psi}_{j}\rangle\ ,
\end{equation}
where $M_{i}=m_{1i}+m_{2i}$.

Integrating in $d^3p$ and multiplying by $|\vec{p}|^{l}$ both sides, we obtain the following equation, written in matrix form
\begin{equation}
\label{eq:wf_cc3}
\int{d^3p}|\vec{p}|^{l}\langle\vec{p}|\tilde{\Psi}\rangle=G\ v\ \int_{k<\Lambda}{d^3k}|\vec{k}|^{l}\langle\vec{k}|\tilde{\Psi}\rangle\ .  
\end{equation}
From Eq. (\ref{eq:wf_cc3}) follows again the condition to find the pole
\begin{equation}
\label{eq:polecondition2}
det(1-Gv)=0\ .
\end{equation}

Eq. (\ref{eq:wf_cc3}) can be rewritten as
\begin{equation}
\label{eq:wf_cc4}
\sum_{j}v_{ij}\ \int{d^3p}|\vec{p}|^{l}\langle\vec{p}|\tilde{\Psi}_{j}\rangle=[G_{i}^{\alpha}]^{-1}\int{d^3p}|\vec{p}|^{l}\langle\vec{p}|\tilde{\Psi}_{i}\rangle\ , 
\end{equation}
which, substituted in Eq. (\ref{eq:wf_cc2}), gives
\begin{equation}
\label{eq:wf_cc5}
\langle\vec{p}|\tilde{\Psi}_{i}\rangle=\frac{|\vec{p}|^{l}\Theta(\Lambda-p)}{E-M_{i}-\frac{\vec{p}^2}{2\mu_{i}}}[G_{i}^{\alpha}]^{-1}\int_{k<\Lambda}{d^3k}|\vec{k}|^{l}\langle\vec{k}|\tilde{\Psi}_{i}\rangle\ . 
\end{equation}

We can now define the partial probability
\begin{equation}
\label{eq:partprob}
P_{i}=\int{d^3p}|\langle\vec{p}|\Psi_{i}\rangle|^2 
\end{equation}
and write the normalization condition for the wave functions
\begin{equation}
\label{eq:prob}
\sum_{i}P_{i}=1\ . 
\end{equation}
Substituting Eq. (\ref{eq:wf_cc}) in the last equation we find the generalization to many channels of Eq. (\ref{eq:prob3})
\begin{equation}
\label{eq:partprob2}
P_{i}=\left[-\frac{dG_i}{dE}\right]_{E=E_{\alpha}}\ \frac{1}{\left[G_{i}^{\alpha}\right]^2}\left|\int_{k<\Lambda}{d^3k}|\vec{k}|^{l}\langle\vec{k}|\tilde{\Psi}_{i}\rangle\right|^2\ . 
\end{equation}

\subsection{Wave functions in coordinate space}
In coordinate space we have
\begin{equation}
\label{eq:wfcoor_cc}
\begin{split}
\langle\vec{x}|\Psi_{i}\rangle&=\int{\frac{d^3p}{(2\pi)^{2/3}}}e^{i\vec{p}\vec{x}}\langle\vec{p}|\Psi_{i}\rangle=\int_{p<\Lambda}{\frac{d^3p}{(2\pi)^{2/3}}}e^{i\vec{p}\vec{x}}(4\pi)^{1/2}\\&\times\sum_{m}a_{m}Y_{lm}(\hat{p})\frac{|\vec{p}|^{l}}{E-M_{i}-\frac{\vec{p}^2}{2\mu_{i}}}\frac{1}{G_{i}^{\alpha}}\int_{k<\Lambda}{d^{3}k}|\vec{k}|^{l}\langle\vec{k}|\tilde{\Psi}_{i}\rangle\ , 
\end{split}
\end{equation}
 
which, expanding the plane wave by means of Eq. (\ref{eq:planewave}), becomes
\begin{equation}
\begin{split}
\label{eq:wfcoor_cc}
\Psi_{i}(\vec{x})=\langle\vec{x}|\Psi_{i}\rangle&=\int_{p<\Lambda}{\frac{d^3p}{(2\pi)^{2/3}}}(4\pi)^{1/2}\sum_{m}a_{m}Y_{lm}(\hat{r})\ i^{l}\ j_{l}(pr)\frac{|\vec{p}|^{l}}{E-M_{i}-\frac{\vec{p}^2}{2\mu_{i}}}\frac{1}{G_{i}^{\alpha}}\\&\times\int_{k<\Lambda}{d^3k}|\vec{k}|^{l}\langle\vec{k}|\tilde{\Psi}_{i}\rangle\ . 
\end{split}
\end{equation}

As in the previous case, we obtain the expression of the wave function at the origin in coordinate space using Eq. (\ref{eq:bessel}):
\begin{equation}
\label{eq:wfcoor_cc}
\Psi_{i}(\vec{x}\equiv 0)=\frac{(4\pi)^{1/2}\sum_{m}a_{m}Y_{lm}(\hat{r})i^{l}|\vec{r}|^{l}}{(2\pi)^{3/2}(2l+1)!!}\int_{k<\Lambda}{d^3k}|\vec{k}|^{l}\langle\vec{k}|\tilde{\Psi}_{i}\rangle\ . 
\end{equation}

Now we go back to Eq. (\ref{eq:wf_cc3}). Defining, as in the previous section
\begin{equation}
\label{eq:psidef}
\hat{\Psi}_{i}\equiv \Psi_{i}(\vec{x}\rightarrow 0)\left[\frac{(4\pi)^{1/2}\sum_{m}a_{m}Y_{lm}(\hat{r})i^{l}|\vec{r}|^{l}}{(2\pi)^{3/2}(2l+1)!!}\right]^{-1}=\int_{k<\Lambda}d^3k|\vec{k}|^{l}\langle\vec{k}|\tilde{\Psi}\rangle\ ,  
\end{equation}
we are allowed to rewrite Eq. (\ref{eq:wf_cc3}) as
\begin{equation}
\label{eq:coor}
\hat{\Psi}=G\ v\ \hat{\Psi}\ . 
\end{equation}
Eq. (\ref{eq:coor}) requires for its solution
\begin{equation}
\label{eq:solution}
det(1-vG)=0\  ,
\end{equation}
which is guaranteed for a bound eigenstate of energy $E_{\alpha}$.
So, this equation can be rewritten as
\begin{equation}
\label{eq:coor2}
\left[G^{\alpha}\right]^{-1}\hat{\Psi}=v\hat{\Psi}\ , 
\end{equation}
which allows us to write the expression of the wave function in momentum space in terms of the wave function at the origin of coordinate space, as
\begin{equation}
\label{eq:coor3}
\langle\vec{p}|\Psi_{i}\rangle=(4\pi)^{1/2}\sum_{m}a_{m}Y_{lm}(\hat{p})\frac{\Theta(\Lambda-p)|\vec{p}|^{l}}{E-M_{i}-\frac{\vec{p}^2}{2\mu_{i}}}\left[G_{i}^{\alpha}\right]^{-1}\hat{\Psi}_{i}\ . 
\end{equation}

From the normalization condition in Eq. (\ref{eq:prob}) follows again
\begin{equation}
\label{eq:sumrule}
\begin{split}
\sum_{i}\langle\Psi_{i}|\Psi_{i}\rangle&=\int{d^3p}\sum_{i}|\langle\vec{p}|\Psi_{i}\rangle|^{2}\\&=-\sum_{i}\left[\frac{dG_{i}}{dE}\right]_{E=E_{\alpha}}\frac{1}{\left[G_{i}^{\alpha}\right]^{2}}\hat{\Psi_{i}}^{2}=1\ .
\end{split}
\end{equation}

\subsection{Couplings}
\label{coup}
Now we want to define the couplings in terms of $\hat{\Psi}_{i}$.

We use the version of Eq. (\ref{eq:lipp-schw2}) for the Lippmann-Schwinger equation. Recalling that close to the pole of the eigenfunction of the Hamiltonian associated to the $E_{\alpha}$, only this state  $|\alpha\rangle$ contributes in the sum over the eigenstates, we find
\begin{equation}
\label{eq:tmatrixbs}
t_{ij}=v_{ij}+\sum_{mn}v_{im}\hat{\Psi}_{m}\frac{1}{E-E_{\alpha}}\hat{\Psi}_{n}v_{nj}\ .
\end{equation}

When we look for the couplings as the residues in the pole of the t-matrix, we obtain
\begin{equation}
\label{eq:coupl_cc}
\begin{split}
g_{i}g_{j}&=\lim_{E\rightarrow E_{\alpha}}(E-E_{\alpha})t_{ij}=\sum_{mn}v_{im}\hat{\Psi}_{m}v_{nj}\hat{\Psi}_{n}\\&=
\left[G_{i}^{-1}\hat{\Psi}_{i}G_{j}^{-1}\hat{\Psi}_{j}\right]_{E=E_{\alpha}}\ ,
\end{split}
\end{equation}
which allows us to write the couplings in terms of the wave function at the origin of coordinate space,

\begin{equation}
\label{eq:coupl_cc2}
g_{i}=\left[G_{i}^{\alpha}\right]^{-1}\hat{\Psi}_{i}\ .
\end{equation}
 
It is also possible to rewrite Eq.  (\ref{eq:sumrule}) as
\begin{equation}
\label{eq:sumrule2}
\sum_{i}g_{i}^{2}\left[\frac{dG_{i}}{dE}\right]_{E=E_{\alpha}}=-1\ ,
\end{equation}
in complete analogy with the case of only one channel.

Each of the terms in Eq. (\ref{eq:sumrule2}) (with opposite sign) gives the probability to find a certain channel in the wave function of the bound states.

\section{Generalization to open channels}     
\label{oc}
Now we want to adapt the formalism to the case of open channels.

We work directly in coupled channels where at least one is open. We take again
\begin{equation}
\label{eq:potential_res}
\langle\vec{p}|V|\vec{p}\ '\rangle\equiv (2l+1)\ v\ \Theta(\Lambda-p)\Theta(\Lambda-p')|\vec{p}|^{l}|\vec{p}\ '|^{l}P_{l}(\cos\theta)\ . 
\end{equation}

\subsection{Lippmann-Schwinger equation for open channels}
In order to create a resonance from the interaction of many channels at a certain energy, we must take a channel which is open at this energy and make the two particles collide, starting from an infinite separation at $t=-\infty$. We call this channel, which is asymptotically the scattering state, channel $1$. 

The equations we have to solve are
\begin{equation}
\label{eq:ls_res}
|\Psi\rangle=|\Phi\rangle +\frac{1}{E-H_{0}}V|\Psi\rangle\ ,
\end{equation}
where
\begin{equation}
|\Psi\rangle=
\begin{Bmatrix}
|\Psi_{1}\rangle \\ |\Psi_{2}\rangle \\\vdots \\|\Psi_{N}\rangle
\end{Bmatrix}\ ,\ \ \ \ \ \ \
|\Phi\rangle=
\begin{Bmatrix}
|\Phi_{1}\rangle \\0\\\vdots \\0
\end{Bmatrix}\ ,
\end{equation}
and $|\Phi\rangle=|\vec{p}\ '\rangle$. Once again $\mu_{i}$ is the reduced mass of the system of total mass $M_{i}=m_{1i}+m_{2i}$.

\subsection{Wave function in momentum space}
We can proceed analogously to the bound states case and write the wave functions in momentum space as
\begin{equation}
\begin{split}
\label{eq:wf_res}
&\langle\vec{p}|\Psi_{1}\rangle-\langle\vec{p}|\Phi_{1}\rangle=(4\pi)^{1/2}\sum_{m}a_{m}Y_{lm}(\hat{p})\frac{|\vec{p}|^{l}\Theta(\Lambda-p)}{E-M_1-\frac{\vec{p}^2}{2\mu_{1}}+i\epsilon}\sum_{j}v_{1j}\int_{k<\Lambda}{d^3k}|\vec{k}|^{l}\langle\vec{k}|\tilde{\Psi}_{j}\rangle\ ,\\&
\langle\vec{p}|\Psi_{i}\rangle=(4\pi)^{1/2}\sum_{m}a_{m}Y_{lm}(\hat{p})\frac{|\vec{p}|^{l}\Theta(\Lambda-p)}{E-M_i-\frac{\vec{p}^2}{2\mu_{i}}+i\epsilon}\sum_{j}v_{ij}\int_{k<\Lambda}{d^3k}|\vec{k}|^{l}\langle\vec{k}|\tilde{\Psi}_{j}\rangle\ ,\ \ \ \ \ \ \ \ \ i\neq 1\ .
\end{split}
\end{equation}
In the bound state case we had $E<M_i$ and $E-M_i-\vec{p}^2/2\mu$ cannot be zero for any value of $E$. We only have descrete eigenstates for some energies. Now, we are dealing with open channels and, since any value of $E$ is allowed and we can have singularities when $E=M_i+\vec{p}^2/2\mu_i$, we need to put $+i\epsilon$ in order to guarantee a solution to the Lippmann-Schwinger equations.  

In order to make the problem technically easy we shall prepare the state $|\Phi_{1}\rangle$ such that it contains only the $l$-wave:
\begin{equation}
\label{eq:phi1}
|\Phi_{1}\rangle=\int{d^3p'}a(\vec{p}\ ')|\vec{p}\ '\rangle\ .
\end{equation}
We can choose $a(\vec{p}\ ')$ such that
\begin{equation}
\label{eq:ap}
a(\vec{p}\ ')=(4\pi)^{1/2}Y_{lm}(\hat{p'})a(p')\ ,
\end{equation}
where $a(p')$ is, for instance, a Gaussian around $p_1$ and $m$ is fixed.

Thus, we find that
\begin{equation}
\label{eq:phi12}
\begin{split}
\langle\vec{p}|\Phi_{1}\rangle&=\int{d^3p'}(4\pi)^{1/2}Y_{lm}(\hat{p}')a(p')\langle\vec{p}|\vec{p}\ '\rangle\\&=\int{d^3p'}(4\pi)^{1/2}Y_{lm}(\hat{p}')\delta^{(3)}(\vec{p}-\vec{p}\ ')a(p')\\&=(4\pi)^{1/2}Y_{lm}(\hat{p})a(p)\ ,
\end{split}
\end{equation}
with $a(p)$ normalized such that $\int{d^3p}\ a(p)|\vec{p}|^{l}=1$.

Now all the terms in $|\Psi_{i}\rangle$ have the same angular dependence and we can write
\begin{equation}
\label{eq:wf_res2}
\begin{split}
&\langle\vec{p}|\tilde{\Psi}_{1}\rangle =a(p)+\frac{\Theta(\Lambda-p)|\vec{p}|^{l}}{E-M_1-\frac{\vec{p}^2}{2\mu_{1}}+i\epsilon}\sum_{j}v_{1j}\int_{k<\Lambda}{d^3k}|\vec{k}|^{l}\langle\vec{k}|\tilde{\Psi}_{j}\rangle\ ,\\& \langle\vec{p}|\tilde{\Psi}_{i}\rangle =\frac{\Theta(\Lambda-p)|\vec{p}|^{l}}{E-M_i-\frac{\vec{p}^2}{2\mu_{i}}+i\epsilon}\sum_{j}v_{ij}\int_{k<\Lambda}{d^3k}|\vec{k}|^{l}\langle\vec{k}|\tilde{\Psi}_{j}\rangle\ , \ \ \ \ \ \ \ \ \ \ \ \ \ \ i\neq 1\ .
\end{split}
\end{equation}

Now again we integrate in $d^3p$ and multiply by $|\vec{p}|^{l}$, and since
\begin{equation}
\int{d^3p}|\vec{p}|^{l}\langle\vec{p}|\tilde{\Phi}_{1}\rangle =1 ,
\end{equation}
we find
\begin{equation}
\label{eq:wf_res3}
\begin{split}
&\int{d^3p}|\vec{p}|^{l}\langle\vec{p}|\tilde{\Psi}_{1}\rangle =1+G_{1}\sum_{j}v_{1j}\int_{k<\Lambda}{d^3k}|\vec{k}|^{l}\langle\vec{k}|\tilde{\Psi}_{j}\rangle\ ,\\&\int{d^3p}|\vec{p}|^{l}\langle\vec{p}|\tilde{\Psi}_{i}\rangle=G_{i}\sum_{j}v_{ij}\int_{k<\Lambda}{d^3k}|\vec{k}|^{l}\langle\vec{k}|\tilde{\Psi}_{j}\rangle\ , \ \ \ \ \ \ \ \ \ \ \ \ \ \ i\neq 1\ , 
\end{split}
\end{equation}
with $G_i$ defined as 
\begin{equation}
\label{eq:loop_res}
G_i=\int_{p<\Lambda}d^{3}p\frac{|\vec{p}|^{2l}}{E-M_i-\frac{\vec{p}^2}{2\mu_{i}}+i\epsilon}\ .
\end{equation}

As in the previous case we can define
\begin{equation}
\label{eq:psihat}
\hat{\Psi}_{i}=\int_{p<\Lambda}{d^3p}|\vec{p}|^{l}\langle\vec{p}|\tilde{\Psi}_{i}\rangle\ ,
\end{equation}
which allows us to rewrite Eqs. (\ref{eq:wf_res3}) as
\begin{equation}
\label{eq:psihat2}
\begin{split}
&\hat{\Psi}_{1}=1+G_{1}\sum_{j}v_{1j}\hat{\Psi}_{j}\ ,\\&
\hat{\Psi}_{i}=G_{i}\sum_{j}v_{ij}\hat{\Psi}_{j}\ , \ \ \ \ \ \ \ \ \ \ \ \ \ \ i\neq 1\ .
\end{split}
\end{equation}

In matrix form, we have
\begin{equation}
\label{eq:psihat3}
(1-Gv)\hat{\Psi}=\begin{Bmatrix}
1 \\ 0 \\\vdots \\0
\end{Bmatrix}
\end{equation}
and hence
\begin{equation}
\label{eq:psihat4}
\hat{\Psi}_{i}=(1-Gv)_{i1}^{-1}\ .
\end{equation}

The $N\times N$ scattering matrix is still given by
\begin{equation}
\label{eq:scatmatrix}
t=(1-vG)^{-1}v=(v^{-1}-G)^{-1}\ ,
\end{equation}
and by means of Eq. (\ref{eq:psihat4}) we can write 
\begin{equation}
\label{eq:psihat5}
\begin{split}
v_{ij}\hat{\Psi}_{j}&=v_{ij}(1-Gv)^{-1}_{j1}\\&=(v^{-1}-G)^{-1}_{i1}=t_{i1}\ .
\end{split}
\end{equation}

Going back to Eq. (\ref{eq:wf_res2}), follows
\begin{equation}
\begin{split}
\label{eq:psihat6}
&\langle\vec{p}|\Psi_{1}\rangle=(4\pi)^{1/2}Y_{lm}(\hat{p})\left(a(p)+\frac{\Theta(\Lambda-p)|\vec{p}|^{l}}{E-M_1-\frac{\vec{p}^2}{2\mu_{1}}+i\epsilon}t_{11}\right)\ ,\\&\langle\vec{p}|\Psi_{i}\rangle=(4\pi)^{1/2}Y_{lm}(\hat{p})\frac{\Theta(\Lambda-p)|\vec{p}|^{l}}{E-M_i-\frac{\vec{p}^2}{2\mu_{i}}+i\epsilon}t_{i1}\ ,\ \ \ \ \ \ \ \ \ \ \ \ \ \ \ \ \ \ \ \ \ \  i\neq 1\ . 
\end{split}
\end{equation}

\subsection{Wave functions in coordinate space}
In coordinate space, the wave functions can be written as
\begin{equation}
\begin{split}
\label{eq:coord_res}
&\langle\vec{x}|\Psi_{1}\rangle=(4\pi)^{1/2}\ i^{l}\ Y_{lm}(\hat{r})\int{\frac{d^3p}{(2\pi)^{3/2}}}\ j_{l}(pr)\left(a(p)+\frac{\Theta(\Lambda-p)|\vec{p}|^{l}}{E-M_1-\frac{\vec{p}^2}{2\mu_{1}}+i\epsilon}t_{11}\right)\ ,\\&
\langle\vec{x}|\Psi_{i}\rangle=(4\pi)^{1/2}\ i^{l}\ Y_{lm}(\hat{r})\int{\frac{d^3p}{(2\pi)^{3/2}}}\ j_{l}(pr)\frac{\Theta(\Lambda-p)|\vec{p}|^{l}}{E-M_i-\frac{\vec{p}^2}{2\mu_{i}}+i\epsilon}t_{i1}\ \ \ \ \ \ \ \ \ \ \ \ \ \ \ \ \  \ \ \ \ \ \ \ \ \ i\neq 1\ .
\end{split}
\end{equation}

We take again the limit of small argument for the Bessel's functions, Eq. (\ref{eq:bessel}), obtaining
\begin{equation}
\label{eq:coord_res2}
\begin{split}
\langle\vec{x}\rightarrow 0|\Psi_{1}\rangle&=\frac{(4\pi)^{1/2}Y_{lm}(\hat{r})i^{l}|\vec{r}|^{l}}{(2\pi)^{3/2}(2l+1)!!}\left[1+G_{1}t_{11}\right]\\&=\frac{(4\pi)^{1/2}Y_{lm}(\hat{r})i^{l}|\vec{r}|^{l}}{(2\pi)^{3/2}(2l+1)!!}\left[1+G_{1}\sum_{j}v_{ij}\hat{\Psi}_{j}\right]\\&=\frac{(4\pi)^{1/2}Y_{lm}(\hat{r})i^{l}|\vec{r}|^{l}}{(2\pi)^{3/2}(2l+1)!!}\hat{\Psi}_{1}\ ,
\end{split}
\end{equation}

and, similarly

\begin{equation}
\label{eq:coord_res3}
\langle\vec{x}\rightarrow 0|\Psi_{i}\rangle=\frac{(4\pi)^{1/2}Y_{lm}(\hat{r})i^{l}|\vec{r}|^{l}}{(2\pi)^{3/2}(2l+1)!!}\hat{\Psi}_{i}\ ,\ \ \ \ \ \ \ \ \ \ \  \ \ \ \ \ \ \ \ \ i\neq 1\ .
\end{equation}

\subsection{Relation between couplings and wave functions at the origin}
In the vicinity of a resonance
\begin{equation}
\label{eq:tmat_res}
t_{ij}\simeq\frac{g_{i}g_{j}}{E-E_{R}+i\frac{\Gamma}{2}}\ .
\end{equation}
Hence
\begin{equation}
\label{eq:coup_res}
\frac{t_{i1}}{t_{11}}=\frac{g_{i}}{g_{1}}=\frac{\hat{\Psi}_{i}G_{i}^{-1}}{\hat{\Psi}_{1}G_{1}^{-1}}\ .
\end{equation}

We can also use Eq. (\ref{eq:lipp-schw2}) for the Lippmann-Schwinger equation and repeat the steps of section (\ref{coup}), and find analogously that
\begin{equation}
\label{eq:coup_res2}
g_{i}=\hat{\Psi}_{i}G_{i}^{-1}\ .
\end{equation}

In the case of resonances we cannot directly derive the sum rule in Eq. (\ref{eq:sumrule2}), since it follows from the normalization condition of the wave function in coordinate space, which is not finite anymore. However, it still holds in the pole in the complex plane (see \cite{jidohyodo} for a different derivation), where again we have
\begin{equation}
\label{eq:sumrule_oc}
\sum_{i}g_{i}^{2}\left[\frac{dG_{i}}{dE}\right]_{E=E_{P}}=-1\ ,
\end{equation}
with $E_{P}$ the position of the complex pole.

\section{The test for the $\rho$ resonance}     
\label{rho}
By means of the sum rule in Eq. (\ref{eq:sumrule_oc}) we can try to find out whether a resonance is created by the interaction of two particles or not. We want to apply this tool to the specific case of the $\rho$ resonance. 

It is known that the $\rho$ is not dynamically generated by the interaction of two $\pi$ mesons, but it is basically a genuine resonance. Thus, we expect the sum rule not to be satisfied.

In order to quantify the statement, we perform two test. First we use a good model based on chiral unitary theory for the $\rho$ meson, and then we propose a pure phenomenological test, where only $\pi\pi$ data are used.

\subsection{Model for the $\rho$ meson}
\label{suba}
We follow here the approach of \cite{palomar} but slightly modified in order to account for the $p$-wave character of the loop function. Hence, we take
\begin{equation}
\label{eq:pot_chen}
v=-\frac{2}{3f^2}\left(1+\frac{2G_V^2}{f^2}\frac{s}{M_{\rho}^{2}-s}\right)\ ,
\end{equation}
where $M_{\rho}$ is the bare $\rho$ mass, $f$ is the $\pi$ decay constant and $G_V$ the coupling to $\pi\pi$ in the formalism of \cite{Gasser:1983yg}, where $G_V \simeq f/\sqrt{2}$.

In \cite{chen} this model was fitted to $\pi\pi$ data in $I=1$ and the values $f=87.4\ MeV$, $G_V=53\ MeV$ and $M_{\rho}=837.3\ MeV$ were obtained (very similar to those used in \cite{palomar}). Here we use the same potential but the factor $p^2$ in $V$ is removed to get the potential $v$ of Eq. (\ref{eq:potenziale}), which does not depend on the momentum. Instead, here, the $p^2$ factor is included in the loop function (see discussion after Eq. (\ref{eq:loop})). This means we have to redo the fit to the data using
\begin{equation}
\label{eq:ampl_test1}
t=\frac{1}{v^{-1}-G}\ ,
\end{equation} 
with 
\begin{equation}
\label{eq:G_test1}
G(s)=\int{\frac{d^3q}{(2\pi)^3}}\frac{q^2}{s-(\omega_{1}+\omega_{2})^2+i\epsilon}\left(\frac{\omega_{1}+\omega_{2}}{2\omega_{1}\omega_{2}}\right)\ ,
\end{equation}
and $\omega_1=\omega_2=\sqrt{m_{\pi}^2+q^2}$, where a relativistic reformulation is assumed \cite{npa}. The loop function of Eq. (\ref{eq:G_test1}) is regularized by means of a cutoff $q_{max}$. The $\pi\pi$ phase shift is then given by (see \cite{palomar, chen})
\begin{equation}
\label{eq:phase}
T=p^2t=\frac{-8\pi\sqrt{s}}{p\cot{\delta(p)}-ip}\ ,
\end{equation}
with $p$ the momentum of the pion. The best fit produces the values
\begin{equation}
\begin{split}
\label{eq:bestfit}
&f=93\ MeV\ , \\&
G_V=53\ MeV\ , \\&
M_{\rho}=855.36\ MeV\ , \\&
q_{max}=661.52\ MeV\ .
\end{split}
\end{equation}
\begin{figure}[h!]
\includegraphics[width=15cm,height=11cm]{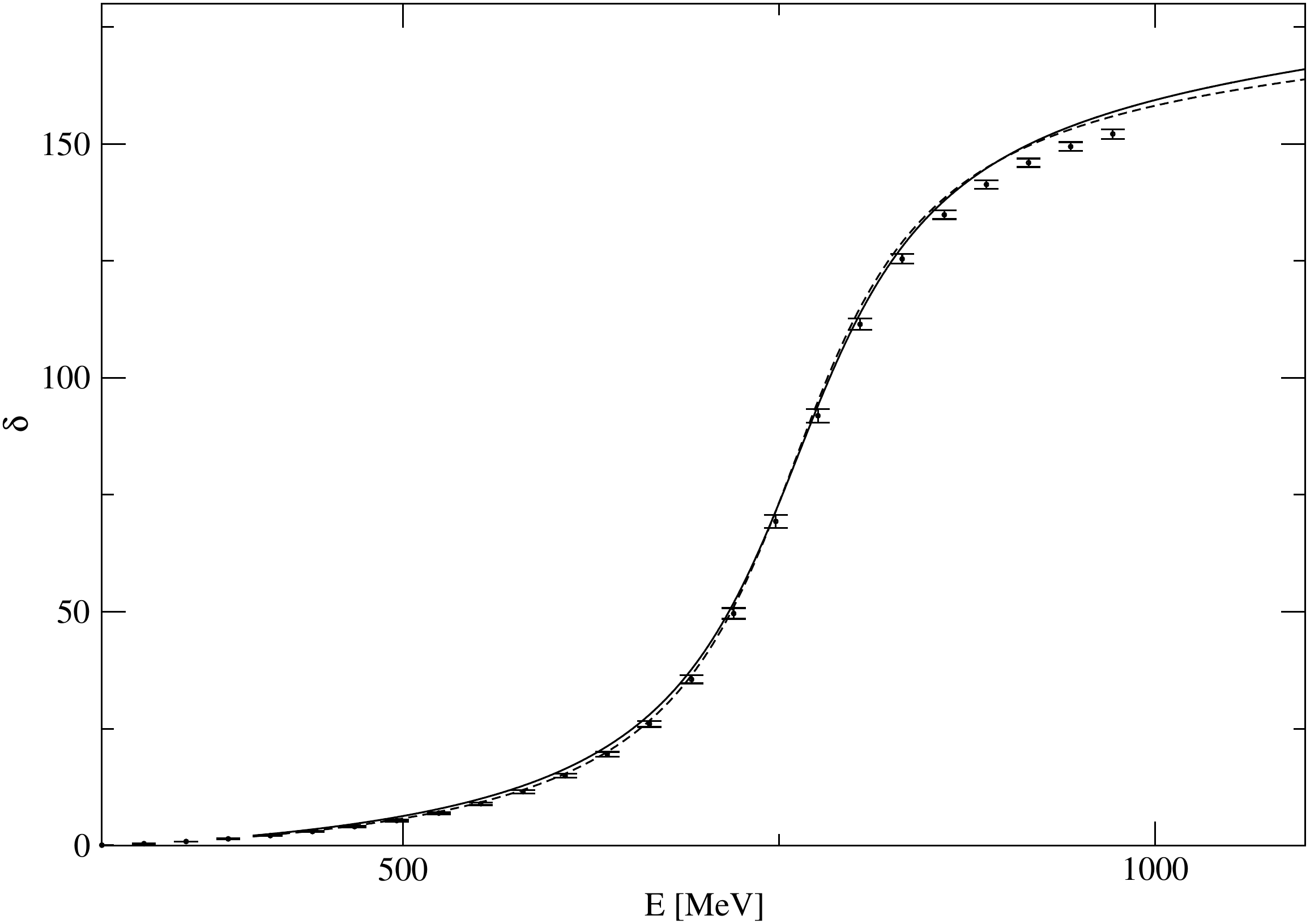}
\caption{The solid curve represents the $\pi\pi$ scattering $p$-wave phase shift obtained in \cite{chen} using the model based on chiral unitary theory \cite{chen}. The dashed curve is the best fit obtained with the new approach. The data are taken from \cite{Garcia}, obtained using the Roy equations.}
\label{fig:fig1}
\end{figure}
The results can be seen in Fig. \ref{fig:fig1}. The results at higher energies could be improved including the $K\bar{K}$ channel and it will be shown in subsection \ref{subc}.
In order to apply the sum rule to the case of a resonance, we need to extrapolate the amplitude to the complex plane and look for the complex pole $s_0$  in the second Riemann sheet. This is done by changing $G$ to $G^{II}$ in Eq. (\ref{eq:ampl_test1}), as will be described below to obtain $t^{II}$.

$G^{II}(s)$ is the analytic continuation to the complex plane of the loop function in $p$-wave for the two pions, 
\begin{equation}
\label{eq:G_rho}
G^{II}(s)=G^{I}(s)+i\frac{p^3}{4\pi\sqrt{s}}\ ,\ \ \ \ \ \ \ \ \ \ \ Im(p)> 0\ ,
\end{equation} 
$G^{I}$ and $G^{II}$ are the loop functions in the first and second Riemann sheet and $G^{I}$ is given by Eq. (\ref{eq:G_test1}).

We are now able to determine the coupling $\tilde{g}_{\rho}$ as the residue in the pole of the amplitude,
\begin{equation}
\label{eq:greal}
\tilde{g}^2_{\rho}=\lim_{s\to s_{0}}(s-s_{0})t^{II}\\ . 
\end{equation}

We can use the sum rule for the single $\pi\pi$ channel in order to evaluate the contribution of this channel to the production of the resonance,
\begin{equation}
\label{eq:sr}
-\tilde{g}^2_{\rho}\left[\frac{dG^{II}(s)}{ds}\right]_{s=s_{0}}=1-Z\ ,
\end{equation}
where $Z$ represents the probability that the $\rho$ is not a $\pi\pi$ molecule but something else.

For the best fit to the data, we find the pole in
\begin{equation}
\label{eq:pole1}
\sqrt{s_0}=(761.70+i\ 71.39)\ MeV\ ,
\end{equation}
while the value of the coupling is
\begin{equation}
\label{eq:coup_test1}
\tilde{g}_{\rho}=(6.86+i\ 0.41)\ ,
\end{equation}
in good agreement with those obtained in \cite{chen}.

Then, we find
\begin{equation}
\label{eq:sum_test1}
\begin{split}
&1-Z=(0.004+i\ 0.267)\ ,\\ &|1-Z|=0.267\ ,
\end{split}
\end{equation}
which indicates that the amount of $\pi\pi$ in the wave function is small. 

\subsection{Phenomenological test}
The exercise done before requires a theoretical model. It would be good to see if it is possible to reach similar conclusions using only data with a pure phenomenological analysis. This is done in this subsection. 

Our aim is to test the sum rule using only experimental data. For $s$-wave there is no problem because the coupling $g$ can be obtained from experiments and $\frac{dG}{dE}$ is a convergent magnitude, even when $q_{max}\to\infty$.
This is, however, not the case for a $p$-wave resonance, such that extra work is required.

The $\rho$ amplitude in a relativistic form can be written as 
\begin{equation}
\label{eq:Trho}
t_{\rho}=\frac{g_{\rho}^2}{s-m_{\rho}^{2} +im_{\rho}\Gamma_{on}\left(\frac{p}{p_{on}}\right)^{3}}\ ,
\end{equation}
where 
\begin{equation}
\label{eq:cmm}
p=\frac{\lambda^{1/2}(s,m^2_{\pi},m^2_{\pi})}{2\sqrt{s}}
\end{equation}
is the three-momentum of the two pions in the center of mass reference frame,
\begin{equation}
\label{eq:qon}
p_{on}=p(\sqrt{s}=m_{\rho})\ ,
\end{equation}
and the coupling is related to the width through the equation
\begin{equation}
\label{eq:gtilde}
g_{\rho}^2=\frac{8\pi m_{\rho}^2\Gamma_{on}}{p_{on}^3}\ .
\end{equation}
The values of the mass $m_{\rho}$ and width $\Gamma_{on}$ of the $\rho$ are given by experiment.

We obtain $t_{\rho}$ in the second Riemann sheet from Eq. (\ref{eq:Trho}) by taking $s$ complex, $s=a+i\ b$, and $p\rightarrow -p$ in the width term. Then we proceed as in the former subsection to get the pole and the coupling. We obtain
\begin{equation}
\label{eq:pole_coup_2}
\begin{split}
&\sqrt{s_0}=(751.13+i\ 68.38)\ MeV\ , \\
&g_{\rho}=(6.58+i\ 1.01)\ ,
\end{split}
\end{equation}
similar to those obtained before.

However, when doing the $1-Z$ test, one does not know which value of the cutoff $q_{max}$ should be used to regularize the $G$ function. Hence, the best one can do is to use natural values of the cutoff and hope that the results are stable for a certain range of $q_{max}$, since $\frac{dG}{ds}$ is only logarithmically divergent.

\begin{table}[ tp ]%
\begin{tabular}{c|c|c}
\hline %
$q_{max}\ [GeV]$ &  $1-Z$ & $|1-Z|$\\\toprule %
$0.6$ & $0.03-i0.26$ & $0.26$ \\
$0.8$ & $0.09-i0.23$ & $0.25$ \\
$1$ & $0.13-i0.21$ & $0.25$ \\
$1.3$ & $0.18-i0.19$ & $0.26$ \\
$1.5$ & $0.20-i0.19$ & $0.28$ \\
$1.7$ & $0.22-i0.18$ & $0.28$ \\
$2$ & $0.24-i0.17$ & $0.29$ \\\hline
\end{tabular}
\caption{Values of $1-Z$ for different cutoffs $q_{max}$.}
\label{tab:vert}\centering %
\end{table}
The values of the strenght $1-Z$ obtained for the $\rho$, changing the cutoff $q_{max}$, are shown in Table \ref{tab:vert}. As we can see, the results are rather stable and similar to those obtained in the former subsection, particularly for $|1-Z|$, with the same conclusion.

Note that since $1-Z$ is a small number, even relatively large uncertainties in this quantity are small errors on $Z$, which measures the amount by which the $\rho$ is not a $\pi\pi$ composite state.

\subsection{Test with two channels}
\label{subc}
In order to get confidence on the conclusions of the previous subsections, we redo the analysis of the $\pi\pi$ data in terms of the $\pi\pi$ and $K\bar{K}$ channels. This was done in \cite{nsd} and \cite{chen} using the transition potentials provided by chiral dynamics. Here we adopt a phenomenological attitude following the procedure used in subsection \ref{suba} for one channel. 

For this purpose we take the potentials
\begin{equation}
\label{eq:potcc}
\begin{split}
&v_{11}=-\frac{2}{3f^2}\left(1+\frac{2G_V^2}{f^2}\frac{s}{M_{\rho}^{2}-s}\right)\ ,\\
&v_{12}=-\frac{2}{3f^2}\alpha\left(1+\beta\frac{2G_V^2}{f^2}\frac{s}{M_{\rho}^{2}-s}\right)\ ,\\
&v_{22}=-\frac{2}{3f^2}\alpha^{'}\left(1+\beta^{'}\frac{2G_V^2}{f^2}\frac{s}{M_{\rho}^{2}-s}\right)\ ,
\end{split}
\end{equation}
where the subscript $1$ is for the $\pi\pi$ channel and $2$ is for $K\bar{K}$. Thus, we keep the structure provided by the chiral Lagrangians, but without adopting the same weights $\alpha$, $\beta$, $\alpha^{'}$ and $\beta^{'}$. From \cite{nsd,chen} one has $\alpha=\sqrt{2}/2$, $\beta=1$, $\alpha^{'}=1/2$ and $\beta^{'}=1$, but here we leave these parameters as free. 

We then evaluate the Bethe-Salpeter equation in coupled channels
\begin{equation}
\label{eq:besal}
t=[1-vG]^{-1}\ v\ ,
\end{equation} 
with $G$ the diagonal matrix $G=diag(G_{\pi\pi}, G_{K\bar{K}})$, where $G_{\pi\pi}$ and $G_{K\bar{K}}$ have the form of Eq. (\ref{eq:G_test1}) and we are now using two different cutoff parameters for each of them, $q_{max}^{\pi\pi}$ and $q_{max}^{K\bar{K}}$ respectively. 

\begin{figure}[h!]
\includegraphics[width=15cm,height=11cm]{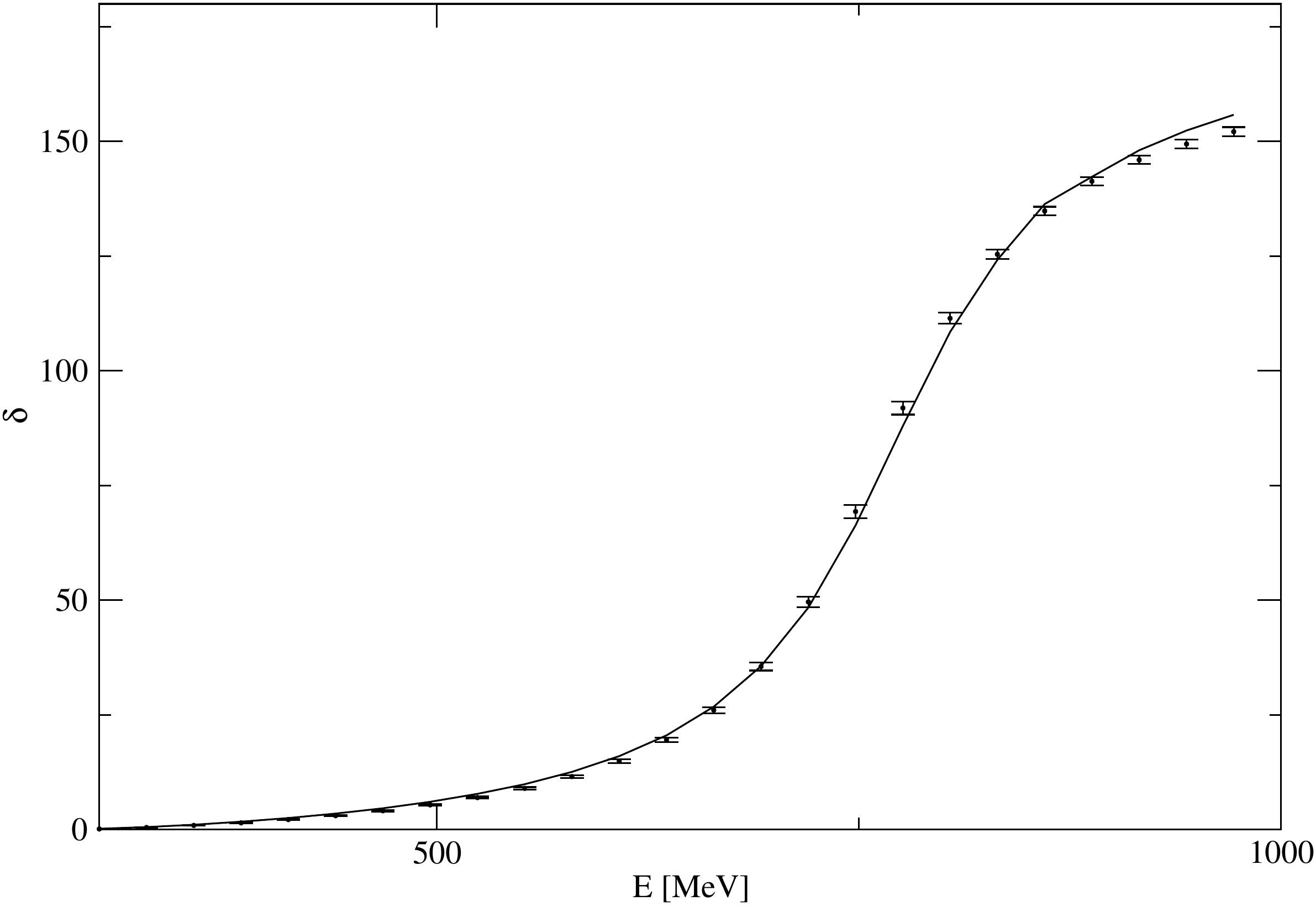}
\caption{The solid curve is the best fit obtained with the new approach. The data are taken from \cite{Garcia}, obtained using the Roy equations.}
\label{fig:fig2}
\end{figure}

Now we carry a $\chi^2$ fit to the data in Fig. \ref{fig:fig1} finding the parameters
\begin{equation}
\begin{split}
\label{eq:bestfitcc}
&f=93.77\ MeV\ , \\&
G_V=52.69\ MeV\ , \\&
M_{\rho}=869.63\ MeV\ , \\&
q^{\pi\pi}_{max}=661.56\ MeV\ , \\&
q^{K\bar{K}}_{max}=610.20\ MeV\ , \\&
\alpha=0.69\ , \\&
\beta=0.81\ , \\&
\alpha^{'}=0.46, \\&
\beta^{'}=0.79\ ,
\end{split}
\end{equation}
which lead to the result shown in Fig. \ref{fig:fig2}. As expected, the data at higher energies improve. 
As we can see, the values obtained for  $\alpha$, $\beta$, $\alpha^{'}$ and $\beta^{'}$ do not differ much from those that would be provided by the chiral Lagrangians.

Now we proceed as before to find the pole and the new couplings, this time for both the $\pi\pi$ and $K\bar{K}$ channels. We obtain
\begin{equation}
\label{eq:pole_cc}
\begin{split}
&\sqrt{s_0}=(768.66+i\ 73.71)\ MeV\ , \\
&g_{\pi\pi}=(6.94+i\ 0.40)\ , \\
&g_{K\bar{K}}=(4.25+i\ 0.01)\ .
\end{split}
\end{equation}
Using these values and  the value of $\frac{dG_i}{ds}$, with $i=\pi\pi, K\bar{K}$, we find
\begin{equation}
\begin{split}
\label{eq:sumcc}
&(1-Z)_{\pi\pi}=-g_{\pi\pi}^2\frac{dG_{\pi\pi}}{ds}=(-5.05\cdot 10^{-5}-i\ 0.27)\ , \\
&(1-Z)_{K\bar{K}}=-g_{K\bar{K}}^2\frac{dG_{K\bar{K}}}{ds}=(1.28\cdot 10^{-2}+i\ 3.32\cdot 10^{-3})\ ,
\end{split}
\end{equation}
from where it follows
\begin{equation}
\begin{split}
\label{eq:modcc}
&|1-Z|_{\pi\pi}= 0.27\ , \\
&|1-Z|_{K\bar{K}}= 0.01 .
\end{split}
\end{equation}
Note that the $K\bar{K}$ channel is not open and hence $G_{K\bar{K}}$ is evaluated in the first Riemann sheet.

We can also evaluate
\begin{equation}
\begin{split}
\label{eq:sumtot}
&1-Z=-g_{\pi\pi}^2\frac{dG_{\pi\pi}}{ds}-g_{K\bar{K}}^2\frac{dG_{K\bar{K}}}{ds}=(-1.28\cdot 10^{-2}-i\ 0.27)\ , \\
&|1-Z|=0.27\ .
\end{split}
\end{equation}
The results in Eq. (\ref{eq:modcc}) lead us to the conclusion that the value of $|1-Z|$ for the $\pi\pi$ channel is essentially unaffected by the inclusion of the $K\bar{K}$ channel. At the same time the exercise has served to  get a feeling of the amount of $K\bar{K}$ present in the $\rho$ meson wave function: we found a negligible fraction of about $1\%$.

Thus, the three tests carried in the work lead us to  conclude that the $\rho$ meson is mostly a dynamical structure with only a small component of $\pi\pi$ in its wave function and a negligible one of $K\bar{K}$.

\section{Conclusions}
We have made an analytical study of the scattering matrix and wave function for the case of the interaction of a pair of hadrons in coupled channels. For this purpose we have followed closely the formalism developed in the chiral unitary approach but using Quantum Mechanics and making all derivations in detail. The study has been done for all partial waves, generalizing work done before for $s$-waves. The study has been done both for bound states and for scattering states. We find novel and interesting relations between the couplings of bound states and resonances to the hadron-hadron channels and the wave function at the origin. Of particular value are the sum rules obtained which allow us to determine the probability to find a certain hadron-hadron component in the wave function. In particular, when the sum of these probabilities is unity we can say that this state is a composite state of hadron-hadron. When the sum differs from unity this difference measures the probability to find a genuine component in the wave function of non hadron-hadron molecular nature. In this sense we extend the rule of compositeness condition derived by Weinberg for $s$-waves, one channel and small binding energies, to any partial wave, several coupled channels, bound states and resonances. As a test we have applied these findings to the rho meson, determining that it is largely a genuine state, with a small component of $\pi \pi$ and one much smaller of $K\bar{K}$, in agreement with other findings based on theoretical studies of the large $N_c$ dependence of the $p$-wave $\pi \pi$ amplitude. 

\section*{Acknowledgments}
 This work is partly supported by DGICYT contract FIS2011-28853-C02-01, the Generalitat Valenciana in the program Prometeo, 2009/090, and 
the EU Integrated Infrastructure Initiative Hadron Physics 3 Project under Grant Agreement no. 283286. 

We would like to thank Y. Zhang and L. R. Dai for checking our numerical results and finding some misprints.

\end{document}